\documentclass[twocolumn,prb,showpacs]{revtex4}

\usepackage{epsfig,amsmath,amssymb,color}
\begin{document}

\title{Spectral function of the $U \rightarrow \infty$ one dimensional Hubbard model at finite temperature and the crossover to the spin incoherent regime}
\author{Mohammad Soltanieh-ha}
\affiliation{Department of Physics, Northeastern University, Boston, Massachusetts 02115, USA}
\author{Adrian E. Feiguin}
\affiliation{Department of Physics, Northeastern University, Boston, Massachusetts 02115, USA}

\date{\today}
\begin{abstract}
The physics of the strongly interacting Hubbard chain (with $t/U \ll 1$) at finite temperatures undergoes a crossover to a spin incoherent regime when the temperature is very small relative to the Fermi energy, but larger than the characteristic spin energy scale. This crossover can be understood by means of Ogata and Shiba's factorized wave function, where charge and spin are totally decoupled, and assuming that the charge remains in the ground state, while the spin is thermally excited and at an effective ``spin temperature''. We use the time-dependent density matrix renormalization group method (tDMRG) to calculate the dynamical contributions of the spin, to reconstruct the single-particle spectral function of the electrons. The crossover is characterized by a redistribution of spectral weight both in frequency and momentum, with an apparent shift by $k_F$ of the minimum of the dispersion.
\end{abstract}


\maketitle

\section{Introduction}

The one-dimensional Hubbard Hamiltonian is a paradigmatic model in condensed matter, not only for its relative simplicity, but mainly because it contains the basic ingredients to understand the physics emerging from strong interactions. Moreover, its higher dimensional counterpart has been assumed for decades to be the minimal model that can explain high temperature superconductivity\cite{ScalapinoReview}, and has acquired even more relevance recently in view of the current efforts to realize it in cold atomic systems\cite{Georgescu2014}. 

The model can be exactly solved by Bethe ansatz, and its low energy physics can be understood in terms of Luttinger liquid (LL) theory\cite{Haldane1981,Gogolin,GiamarchiBook}. 
In a Luttinger liquid, the natural
excitations are collective density fluctuations, that carry either spin
(``spinons''), or charge (``holons''). This leads to the spin-charge separation picture, in
which a fermion injected into the system breaks down into excitations carrying different quantum numbers, each with a characteristic energy scale and velocity (one for the charge, one for the spin).

The phenomenon of spin and charge separation is an important problem in strongly correlated systems, and has intrigued physicists for decades.
Even though these concepts are well established, we keep finding new surprises in unexplored parameter ranges that escaped previous scrutiny. Recently, a previously overlooked regime at finite temperature has come to
light: the ``spin-incoherent Luttinger liquid'' (SILL) \cite{Matveev2004,Fiete2004,Cheianov2004,Cheianov2005,Ejima2006,Fiete2007b,Halperin2007}.
If the spinon bandwidth is much smaller than the holon bandwidth, a small temperature relative to the Fermi energy may actually be felt as a very large temperature by the spins. In fact, the charge will remain very close to the {\it charge} ground-state, but the spins will become totally incoherent, effectively at infinite temperature.
 This regime is characterized by universal properties in the transport, tunneling density of states, and the spectral functions \cite{Fiete2007b}.

This crossover from spin-coherent to spin-incoherent is characterized by a transfer of spectral weight. Remarkably, the photoemission spectrum of the SILL can be understood by assuming that after the spin is thermalized, the charge becomes spinless, with a shift of the Fermi momentum from $k_F$ to $2k_F$. \cite{Feiguin2009b}. Interestingly, it was also shown that a coupling to a spin bath can have a similar effect as temperature, but only in the ground-state.\cite{Feiguin2011,Soltaniehha2012}.
Clearly, this behavior defies our intuition: since the excitation spectrum of the problem is completely determined by the Hamiltonian, at finite temperature (and in the absence of phase transitions) we only expect a redistribution of spectral weight. This paradox can be reconciled in the context of spin-charge separation. If we assume that spin and charge degrees of freedom are completely decoupled, the single-particle spectral function of the electrons can be understood as the convolution of the charge and spin dynamical correlation functions. 

In this paper, we use Ogata and Shiba's factorized wave function\cite{Ogata1990}, complemented by tDMRG\cite{White2004,Daley2004,Feiguin2011vietri,Feiguin2013a} calculations at finite spin temperature, to obtain the spectral function of strongly interacting Hubbard chains ($U\rightarrow\infty$) in the crossover between spin-coherent and spin-incoherent regimes. 

\section{Method}
We study the one-dimensional Hubbard model, which is given by the usual Hamiltonian:
\begin{equation}
H=-t\sum_{i,\sigma} \left(c_{i\sigma}^\dagger c_{i+1,\sigma} +\mathrm{h.c.} \right) + U\sum_i n_{i\uparrow}n_{i\downarrow}
\label{Hubbard}
\end{equation}
where $c_{i\sigma}$ describes a fermionic annihilation operator on site $i$ with spin $\sigma$, $n_{i\sigma}$ is the number operator, and electrons pay an on-site Coulomb repulsion $U>0$.
In the following, all energies are parametrized by the hopping amplitude $t$.

In order to calculate the spectral functions, we follow the formalism developed in Refs.\onlinecite{Sorella1991,Pruschke1991,penc1995spectral,penc1996shadow,Favand1996,Penc1997,Penc1997b}, described in great detail in Ref.\onlinecite{Penc1997}. We sketch the main ideas, avoiding technicalities and directing the reader to the aforementioned references. In the $U\rightarrow\infty$ limit, using Bethe ansatz, Ogata and Shiba demonstrated that the exact solution can be factorized and split into two parts:

\begin{equation}
|\psi^{N,GS}_L\rangle= | \phi\rangle\otimes | \chi\rangle,
\label{gs}
\end{equation}

where $|\phi\rangle$ describes the charge and is comprised by spinless fermionic degrees of freedom, and $|\chi\rangle$ is consistent with a ``squeezed'' chain of $N$ spins, where all the empty sites have been removed. $|\phi\rangle$ is simply the ground-state of a one-dimensional tight-binding chain of $N$ non-interacting spinless fermions on a lattice with $L$ sites. The factorization applies also to excited states as:

\begin{equation}
|\mathrm{\psi(P)}\rangle= | \phi^N_{L,Q}(P,I)\rangle \otimes | \chi^{N_\downarrow}_N(Q,M)\rangle,
\label{ex_state}
\end{equation} 
where $I$ labels a combination of $N$ wave-vectors $k_i L = 2 \pi i + Q$, with $i=-L/2,-L/2+1,\cdots,L/2-1$, that are compatible with the total fermionic momentum $P$. The index $M$ labels all possible configurations of momenta compatible with the total momentum of the spin wave function $Q=2\pi j/N$, with $j=0,1,\cdots,N-1$.
The fermionic part stays coupled to the spin part only by a phase $Q$ introduced at the boundaries, resulting in twisted boundary conditions for the fermions, which ensures momentum conservation for the original problem. This phase is $Q=\pi$ for the ground-state $|\psi^{N,GS}_L\rangle$, and we would fill up the Fermi sea by minimizing the energy with the combination of momenta $\{-\frac{N}{2},-\frac{N}{2}+1,...,\frac{N}{2}-1\}$ \cite{Penc1997b}.

The dynamics of the spinless fermions is governed by a tight-binding Hamiltonian, with an energy dispersion $\epsilon(k)=-2t\cos(k)$, while the spin degrees of freedom are in principle non-interacting. For finite but large $U\gg t$, the physics of the spins can be approximated by that of a Heisenberg spin chain, with $J =4nt^2[1-\sin{(2\pi n)}/(2\pi n)]/U$, where $n=N/L$ is the density \cite{Penc1997,xiang1992charge}.

In order to calculate the Green's functions for the electrons, we start by noticing that destroying an electron in the original Hubbard chain would translate into the annihilation of a fermion in the charge wave function, and a spin in the spin chain (and the opposite effect for the creation operator). This is achieved by introducing new operators such that
\begin{equation}
\label{c_}
c_{i,\sigma}=f_iZ_{l(i),\sigma},
\end{equation}
where $f_i$ is a fermionic annihilation operator without spin, acting at position $i$ on the fermionic wave function, and $Z_{l(i),\sigma}$ removes a spin $\sigma$ at position $l(i)$ from the spin chain. The operator $Z$ has a peculiar behavior, since it destroys the spin and the site itself, making the spin chain $N-1$ sites long. The index function $l(i)$ indicates the position of the spin that belongs to site $i$ of the original chain, after squeezing the wave function and removing all the holes.

The zero-temperature one-particle spectral function is obtained from the imaginary part of the Green's function. We focus only on the contribution from the occupied levels, corresponding to the photoemission spectrum.  In the Lehmann representation, we write it as

\begin{equation}
B(k,\omega)= \sum_{f,\sigma} |\langle f,N-1 | c_{k,\sigma} | GS,N \rangle |^2 \delta (\omega - E_{GS}^N + E_f ^{N-1}).
\label{B_ck}
\end{equation}

Here $c_{k,\sigma}$ destroys an electron with momentum $k$ and spin $\sigma$, $f$ is the final state with $N-1$ particles and $N$ is the total number of electrons. In order to use the factorized wave-function in the calculation, and working in the real space for more convenience, one can re-write this expression as

\begin{eqnarray}
B(k,\omega) = \sum_{f,\sigma} L |\langle f,N-1 | c_{0,\sigma} | GS,N \rangle |^2   \\
\times\delta (\omega - E_{GS}^N + E_f ^{N-1})  \delta_{k,P_{GS}^N-P_f^{N-1}} , \nonumber
\label{B_c0}
\end{eqnarray}
where we have imposed momentum conservation with the term $\delta_{k,P_{GS}^N-P_f^{N-1}}$ and used the definition $c_{j,{\sigma}}=\frac{1}{\sqrt{L}} \sum_{k'} e^{i k' j} c_{k',\sigma}$.  

By taking advantage of the factorized wave-function, Eq.(\ref{ex_state}), and the separated spin and charge operators, Eq.(\ref{c_}) with $i=0$, we obtain the following expression


\begin{equation}
B(k,\omega)= \sum_{Q,\omega',\sigma} D_{\sigma}(Q,\omega') B_Q(k,\omega-\omega').
\label{B_conv}
\end{equation}

In this expression, $B_Q(k,\omega)$ is determined by the spinless fermion operator $f$ as 

\begin{eqnarray}
\label{B_Q}
B_Q(k,\omega)= L \sum_{I} |\langle \phi^{N-1}_{L,Q}(I) | f_0 | \phi^{N,GS}_{L,\pi}\rangle | ^2 \\
\times \delta (\omega - E_{GS,ch}^N + E_{f,ch} ^{N-1}) \times \delta_{k,P_{GS}^N-P_f^{N-1}}, \nonumber
\end{eqnarray}
where the the fermionic energies are obtained from the tight-binding dispersion.
$D_{\sigma}(Q,\omega)$ is defined by the action of the spin operators $Z$: 

\begin{eqnarray}
\label{D_sigma_Q}
D_{\sigma}(Q,\omega)&=& \sum_{M} |\langle \chi_{N-1}(Q,M) |  {Z}_{0,\sigma} | \chi_N^{GS} \rangle | ^2 \\
&&\times\delta (\omega - E_{GS,s}^N + E_{f,s} ^{N-1}). \nonumber
\end{eqnarray}
To calculate $D_{\sigma}(Q,\omega)$, in principle one would need to solve the spin-$\frac{1}{2}$ Heisenberg Hamiltonian to obtain the energies and the wave-functions for $N$ and $N-1$ sites. In the $U\rightarrow \infty$ or $J \rightarrow 0$ limit the excitation spectrum collapses and $D_{\sigma}(Q,\omega)=D_{\sigma}(Q)\delta(\omega)$. The spin transfer structure factor $D_\sigma(Q)$ can be derived by using the spin transfer function $\omega_{j' \rightarrow j,\sigma}$ defined by Ogata and Shiba \cite{Ogata1990,Ogata1991}, and by Sorella and Parola \cite{Sorella1991}. This function gives the amplitude of moving the spin $\sigma$ from site $j'$, to $j$. For $j'=0$ it is given by

\begin{figure}
\includegraphics[width=7cm]{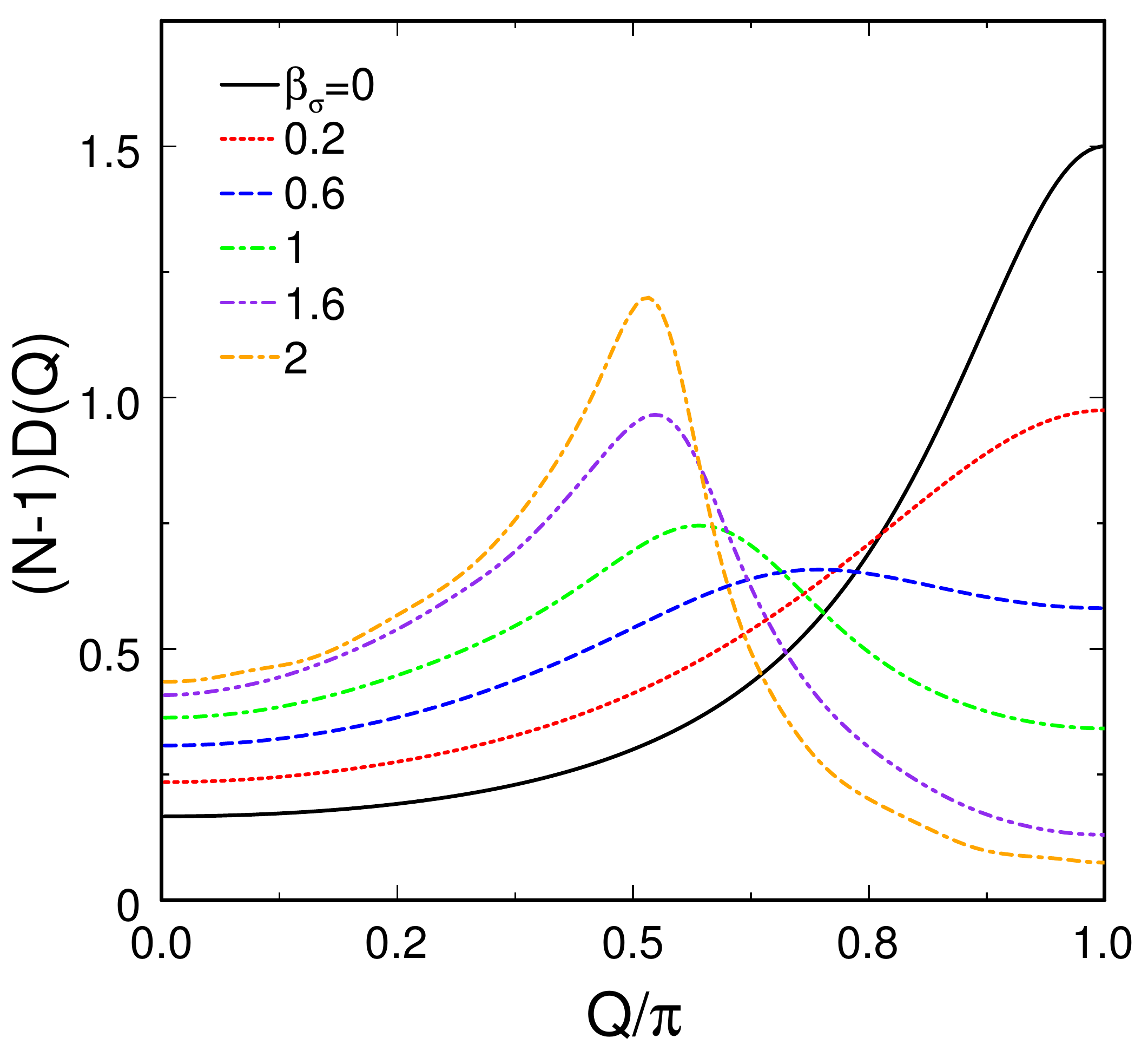}
\caption{ 
Spin transfer structure factor $D_\sigma(Q)$ calculated on Heisenberg spin chain ($J=1$), for different temperatures, obtained with the time-dependent DMRG.
}
\label{fig:zk}
\end{figure}

\begin{equation}
        \omega_{0 \rightarrow j,\sigma} = \langle \chi_N^{GS} \mid P_{j,j-1} \dotsb  P_{1,0} \delta_{\sigma,S_0^Z} \mid \chi_N^{GS} \rangle,
\label{omega}
\end{equation}
where we introduced the permutation operator $P_{j,j-1} = 2 \vec{S}_j \cdot \vec{S}_{j-1} + \frac{1}{2}$, that exchanges the spins at sites $j$ and $j-1$. Using this spin transfer function we get

\begin{equation}
        D_{\sigma}(Q) = \frac{1}{N-1} \sum_{j=0}^{N-2} e^{i (Q+\pi)j} \omega_{0 \rightarrow j,\sigma}.
\label{D_Q}
\end{equation}

For finite but small $J$ one can use the approximation \cite{Penc1997b}

\begin{equation}
D_{\sigma}(Q,\omega) = D_{\sigma}(Q) \delta (\omega - E_s + \epsilon_Q),
\label{cloizeaux}
\end{equation}
where $\epsilon_Q = \frac{\pi}{2} J |\sin(Q-\pi/2)|$ is the des Cloizeaux-Pearson dispersion, and $E_s = - J \ln 2$ is the ground state energy, which are obtained from the exact Bethe ansatz solution for the Heisenberg chain \cite{Cloizeaux}.

This expression for $D_\sigma(Q)$ can be generalized to finite {\it spin temperatures}

\begin{equation}
	D_{\sigma}(Q,\beta_\sigma) = \frac{1}{N-1} \frac{1}{Z_s} \sum_{M} \sum_{j=0}^{N-2} \omega_{0 \rightarrow j,\sigma} e^{i (Q+\pi)j} e^{- \beta_\sigma E_{M,s}},
\label{D_Q_beta}
\end{equation}
where $Z_s$ is the spin partition function. In the following, we express the spin temperature in units of $J$, and the inverse temperature as $\beta_\sigma=1/T_\sigma$.

In order to obtain this quantity numerically, we resort to tDMRG calculations at finite temperature\cite{Feiguin2005a}. The problem reduces to evaluating the thermal average of $\omega_{0 \rightarrow j,\sigma}$, described by Eq.\ref{omega}, for a one-dimensional Heisenberg chain. The antiferromagnetic exchange is set to $J=1$, and the temperature $T_\sigma=1/\beta_\sigma$ is defined in units of $J$. For the final calculation, we approximate $D_\sigma(Q,\omega,\beta_\sigma)$ using Eq.(\ref{cloizeaux}). 

\begin{figure}
\includegraphics[width=8cm]{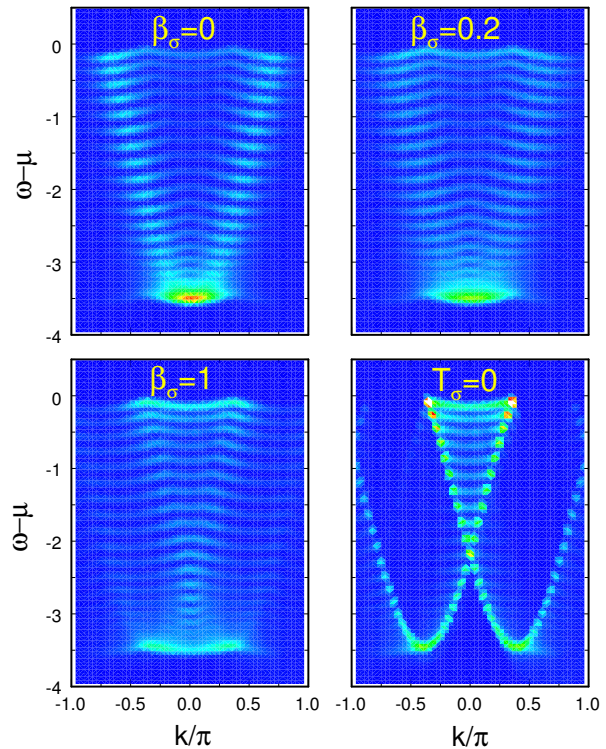}
\caption{ 
Spectral function obtained using the factorized wave function, for $J=0.05$, $L=56$ sites, and $N=42$ fermions. The panels correspond to different values of the ``spin temperature'' (see text).  
}
\label{fig:results}
\end{figure}

Since the Heisenberg chain is weakly entangled at high temperatures, $T_\sigma > J$, we are able to simulate systems as large as 300 sites. We point out that the calculation of $\omega_{0 \rightarrow m,\sigma}$ is not that trivial, since it involves building a string of permutations between all pairs of spins between sites $0$ and $m$. In order to carry it out, we use a recipe similar to the one used to time-evolve the wave-function in tDMRG. Since the structure of the DMRG block decimation always leaves two individual single sites in the original spin basis, we apply the permutation operator between these two sites. We propagate the wave function to the next site, and apply the next permutation. This builds a chain of permutations as the DMRG algorithm sweeps through the lattice. In order to calculate the quantum mechanical average, we need to use two target states, the original thermal state (or the ground state at $T=0$), and the state obtained after applying the permutations. The calculation is done more efficiently if one starts from the middle of the chain. The application of $P_{1,0}$ yields the first average $\omega_{0\rightarrow 1}$. In the next left-to-right iteration we apply $P_{2,1}$ to the previous state, leading to $\omega_{0\rightarrow 2}$. We can repeat this procedure all the way until we reach the right end of the chain, and a single DMRG sweep yields all the correlations needed to construct $D_\sigma(Q,\beta_\sigma)$. 

\begin{figure}
\includegraphics[width=8.5cm]{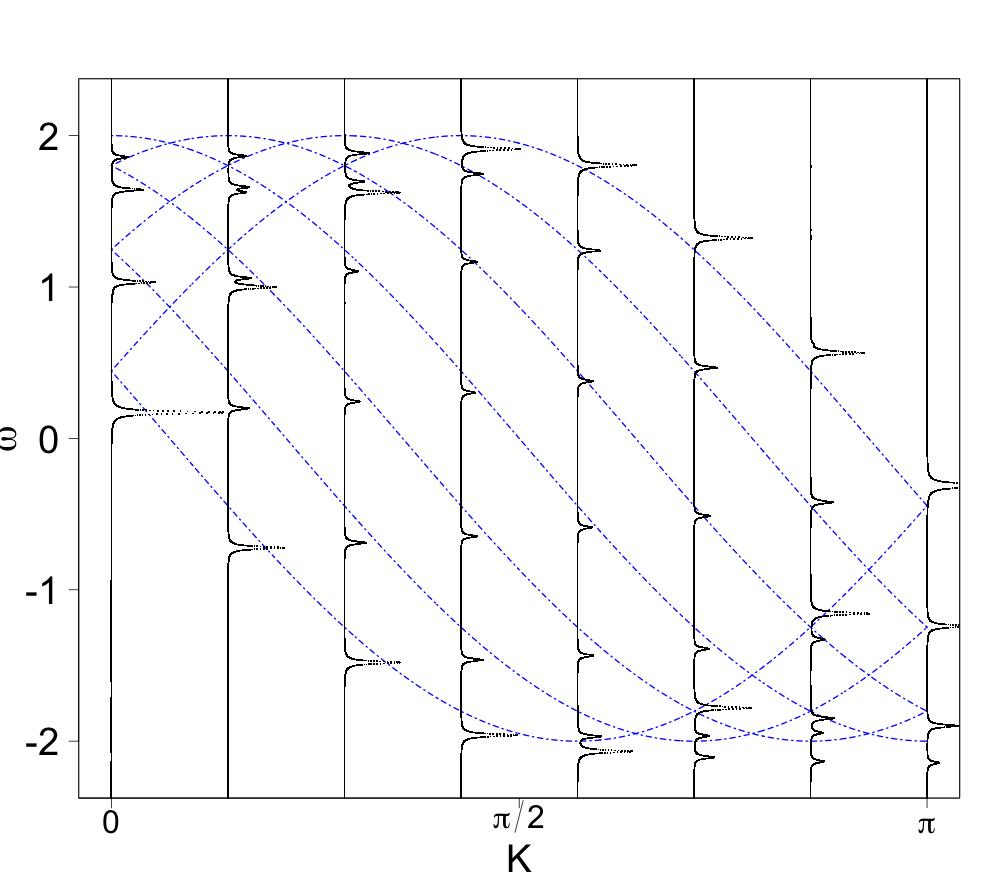}
\caption{ 
Spectral function for a finite chain with $L=14$ sites at half-filling, with $J=0.05$, obtained with exact diagonalization.
}
\label{fig:ED}
\end{figure}

\section{Results}
The results for $D_\sigma(Q,\beta_\sigma)$ are shown in Fig.\ref{fig:zk}, and for the $\beta_\sigma=0$ limit, they are exact, and can be compared to the expression 

\[D_\sigma(Q,\beta_\sigma=0)=\frac{1}{N-1}\frac{3}{10-8\cos{(Q-\pi)}} \]

As the temperature decreases the peak in $Q=0$ shifts toward $Q=\pi/2$, as expected for an antiferromagnetic spin chain. At very low temperatures the correlation function develops a singularity at $Q$, which introduces serious finite-size effects in our calculations. We therefore restrict our simulations to values of $\beta_\sigma$ that we can trust. For $T=0$ we use a fit to the values obtained with exact diagonalization.

In Fig.\ref{fig:results} we show the results for the electronic spectral function after the convolution, using the data for $D_\sigma(Q)$ obtained with tDMRG. We chose the value of $J=0.05$ to compare with the results from Ref.\onlinecite{Feiguin2009b}. We observe identical behavior, the spectral weight shifting in momentum, with the minimum of the dispersion moving from $k=0$ at infinite spin temperature to $k=2k_F$ at zero temperature. In addition, we notice a series of discrete flat energy levels. These energy levels correspond to the convolution of a single charge excitation with $D_\sigma(Q)$. The spacing between them is inversely proportional to the system size $\approx 2w/L$, where $w=4$ is the bandwidth, and it is a finite size effect. This naturally explains the DMRG spectra reported in Ref.\onlinecite{Feiguin2009b}. 

To fully understand this effect, we look at the spectrum of a finite $t-J$ chain using exact diagonalization. We follow the reasoning sketched in Ref.\onlinecite{eder1997}, which at the time was based on pure intuition, and was later proved to be the correct interpretation. It is easy to observe in Fig.\ref{fig:ED} that one could fix a spinon momentum $Q$, and draw a shifted tight-binding dispersion to trace a full holon band. This principle can be used to build the full set of branches. The figure shows the distribution of spectral weights at $T=0$. When the temperature is increased, the spectral weight of the spinons is redistributed according to the spin transfer function $D_\sigma(Q)$, leading to the spinless-like dispersion. 


\section{Conclusions}

We have generalized a formalism originally developed in Refs.\onlinecite{Sorella1991,Pruschke1991,Penc1997,Penc1997b} to study the crossover from the spin-coherent to SILL regime in the spectral signatures of strongly correlated Hubbard chains. The main idea consists of convolving the spectral functions of spinless fermions at zero temperature, and a spin chain at finite temperatures, following the Bethe ansatz prescription. The finite-temperature spin correlation function can be obtained with high-accuracy using tDMRG. The theory reproduces and explains previous numerical results in a very natural way. Although the spectrum is completely determined by the Hamiltonian, the spectral weight of the spinons shifts to high momentum as the temperature increases. This translates into an apparent shift of the minimum of the electronic spectrum by $k_F$. Remarkably, in the SILL regime, the spectral function resembles that of spinless fermions at zero temperature, with a Fermi momentum $2k_F$. Besides a conventional thermal broadening of the line shape, one would also expect to see a broadening in momentum, due to the spread of $D_\sigma(Q)$. This transfer of spectral weight does not mean that the spectrum itself changes with temperature, and it is a clear departure from the non-interacting, or Fermi-liquid picture. Even when instruments may not be able to resolve the fine details of the spectrum, this shift and momentum broadening are a distinctive signature of spin-charge separation that should be experimentally observable.  

\section{Acknowledgement}
The authors acknowledge NSF support through grant DMR-1339564.


\end{document}